\documentclass{Interspeech2024}

\interspeechcameraready

\title{SecureSpectra: Safeguarding Digital Identity from\\Deep Fake Threats via Intelligent Signatures}

\name[affiliation={1}]{Oguzhan}{Baser}
\name[affiliation={2}]{Kaan}{Kale}
\name[affiliation={1}]{Sandeep P.}{Chinchali}

\address{
  $^1$Department of Electrical and Computer Engineering, The University of Texas at Austin, USA\\
  $^2$Department of Electrical and Electronics Engineering, Bogazici University, Turkey}
\email{oguzhanbaser@utexas.edu}


\begin{document}

\maketitle

\begin{abstract}
    
    Advancements in DeepFake (DF) audio models pose a significant threat to voice authentication systems, leading to unauthorized access and the spread of misinformation. We introduce a defense mechanism, SecureSpectra, addressing DF threats by embedding orthogonal, irreversible signatures within audio. SecureSpectra leverages the inability of DF models to replicate high-frequency content, which we empirically identify across diverse datasets and DF models. Integrating differential privacy into the pipeline protects signatures from reverse engineering and strikes a delicate balance between enhanced security and minimal performance compromises. Our evaluations on Mozilla Common Voice, LibriSpeech, and VoxCeleb datasets showcase SecureSpectra's superior performance, outperforming recent works by up to 71\% in detection accuracy. We open-source SecureSpectra to benefit the research community. 
\end{abstract}
\section{Introduction}

The escalating sophistication of DeepFake (DF) technologies is increasingly compromising the security of voice-authenticated applications. Recent DF models can clone voices from recordings as brief as 10-second recordings \cite{coqui}, amplifying the risks in critical domains, such as access to banking and medical records by voice \cite{mustak2023deepfakes}. Consequently, there is an urgent need to safeguard against voice misuse and to ensure the security of voice-based interactions. Notably, our work is partly motivated by DF attacks targeting political leaders \cite{pawelec2022deepfakes, kirchengast2020deepfakes}, which raises concerns similar to the Cambridge Analytica case \cite{pavlikova2021propaganda} and leads to incompliance with GDPR \cite{GDPR}.

Existing state-of-the-art methods use machine learning (ML) models to identify cloned audio \cite{liu23v_interspeech}. However, these ML classifiers struggle against DF models. This is largely because advanced cloning schemes \cite{coqui, openvoice, whisperspeech} rely on Generative Adversarial Networks (GANs), and GANs are optimized to deceive their internal ML classifiers (discriminators) to refine clones. Due to a lack of additional orthogonal information, traditional ML models are not effective in DF detection and upper-bounded by the discriminator performance. Our key innovation lies in enriching the original audio with orthogonal information, which we call \textit{the signature}, to improve the DF detection performance. Also, this method applies to any type of DF model, such as diffusion models \cite{ho2020denoising} and VAEs \cite{kingma2013auto}, in addition to GANs.

\begin{figure}[t]
    \centering
    \includegraphics[width=0.9\linewidth]{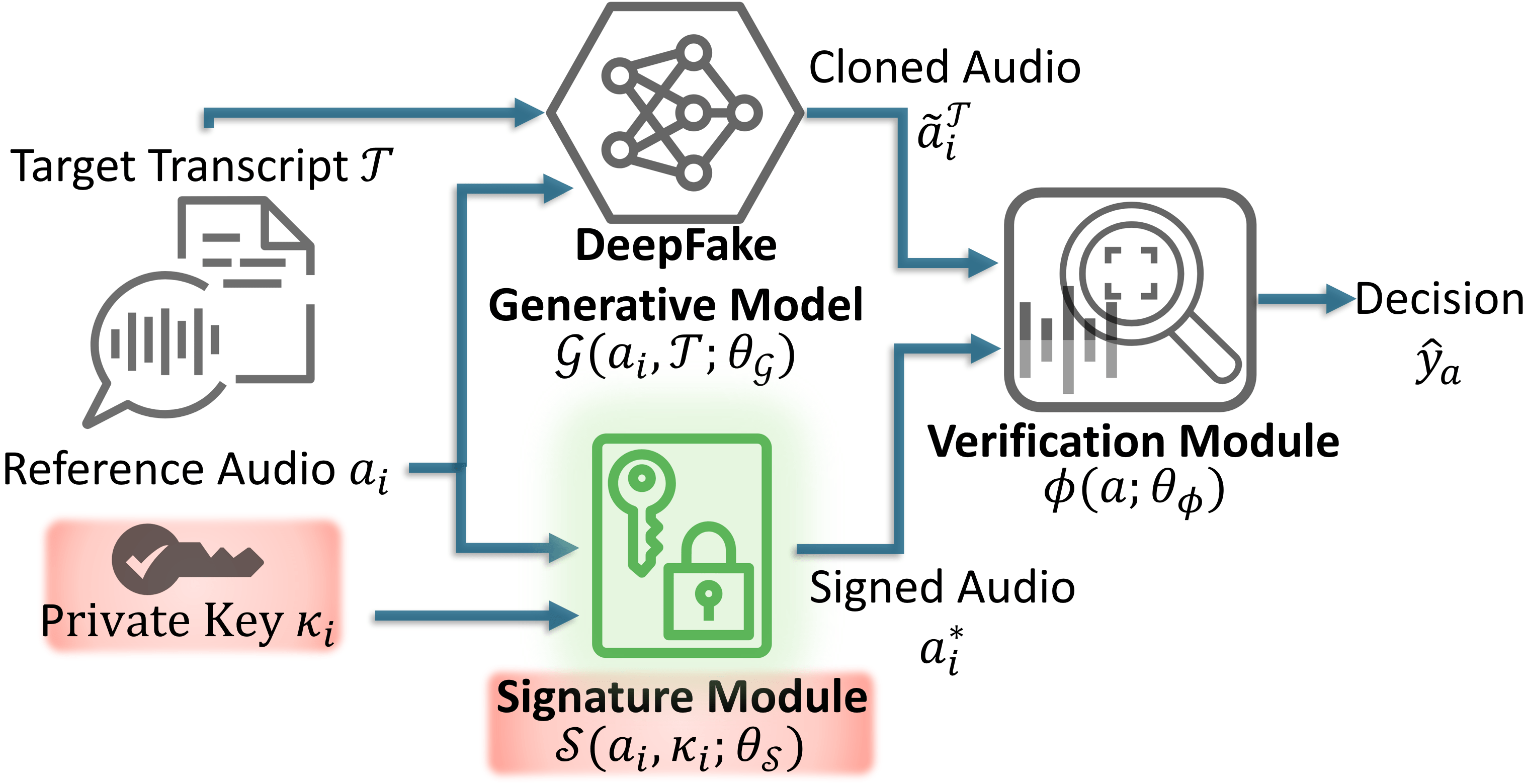}
    \caption{\textbf{Digital Identity Secured Voice Authentication}: Imagine a public figure, Alice, who releases a speech of $a_i$. Malicious Eve (top row) employs a DF model $\mathcal{G}$, parameterized by $\theta_\mathcal{G}$, to mimic Alice's voice in a transcript $\mathcal{T}$, creating a clone $\Tilde{a}_i^\mathcal{T}$. We aim to develop a verification module $\phi$ parameterized by $\theta_\phi$ to decide $\hat{y}_a$ if an audio $a$ comes from Alice. Our approach (bottom row) first gives Alice a private key $\kappa_i$. Then, our novel signature module (green) combines her voice $a_i$ with her key $\kappa_i$ to produce signed audio $a_i^*$. The signed audio closely resembles the original while being distinguishable from fake versions. The verifier $\phi$ can identify the signature in an audio without revealing it. If Eve attempts to use the signed audio $a_i^*$ in her model, the generated clone $\Tilde{a}_i^\mathcal{T}$ does not contain the signature. The key and the signature module are kept confidential (red) to prevent attacks.}
    \label{fig:pipeline}
\end{figure}
\begin{figure}[!h]
\centering
\includegraphics[width=0.85\linewidth]{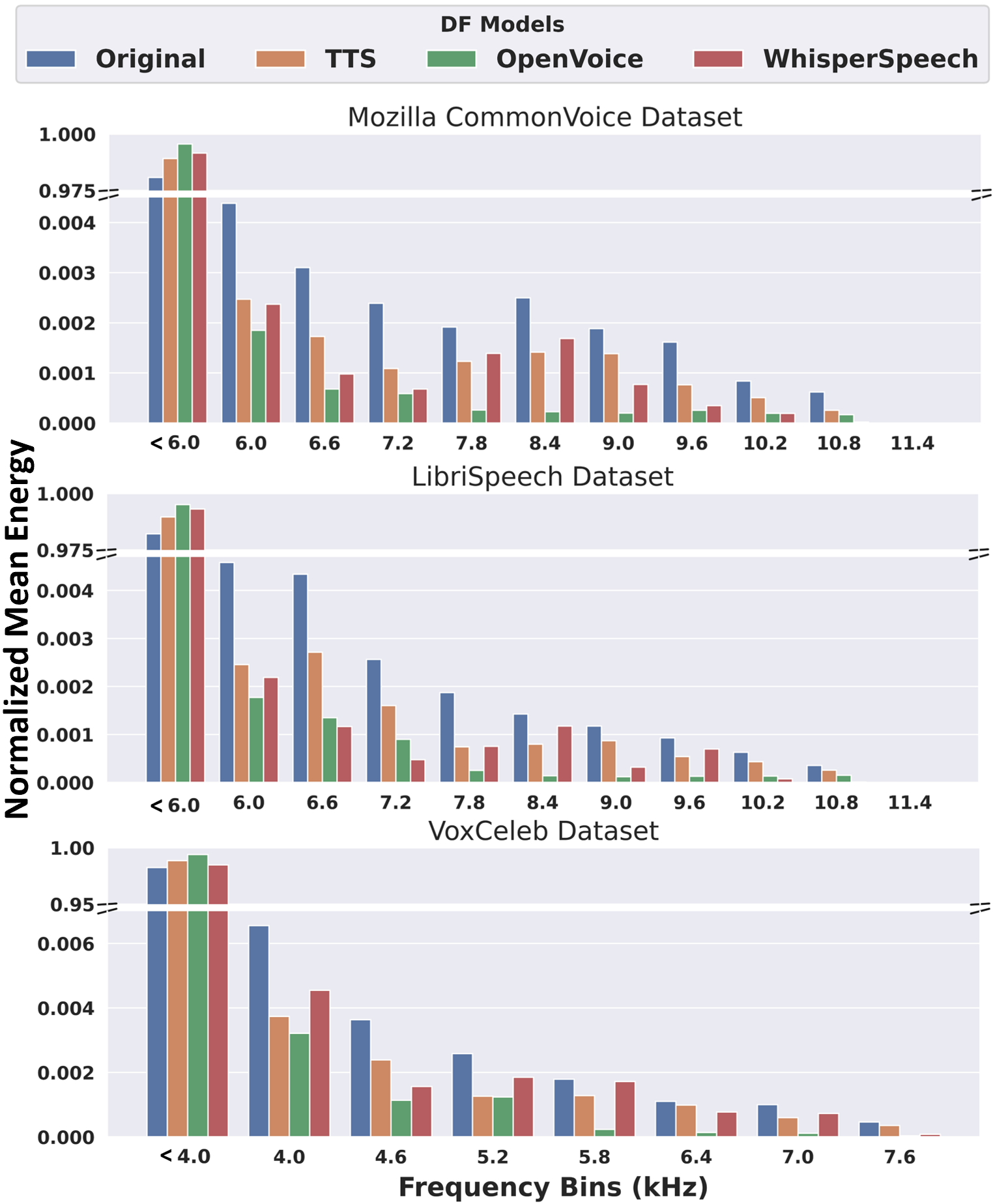}
    \caption{\textbf{Spectral Analysis of Original and Cloned Audio:} We empirically analyzed the spectral content across the original audio recordings (blue) and their cloned counterparts (orange, green, red) generated by state-of-the-art DF models. The analysis encompasses CommonVoice, LibriSpeech, and VoxCeleb datasets, with all audio samples converted into spectrograms. The frequency spectrum was segmented into bins, each representing a 600 Hz bandwidth, where the energy content within each bin was averaged and plotted. The results, derived from testing on three known audio datasets with three advanced DF models, highlight a \textbf{discernible attenuation in the HF components in the DF-generated audio compared to the original ones}, indicating a characteristic shortfall of the DF models in replicating the HF energy profile of genuine audio recordings.}
    \label{fig:empirical_result}
\end{figure}

Specifically, as shown in Fig.\ref{fig:pipeline}, our design\footnote{\url{https://github.com/UTAustin-SwarmLab/SecureSpectra} \label{repourl}} provisions an irreversible signal processing module (green) at the audio owner's end. This module enables authenticated users to sign the audio with their private key, generating signed audio that is resistant to the extraction of the exact signature without the private key yet remains verifiable. Our key technical insight, as we empirically demonstrate in Fig.\ref{fig:empirical_result}, lies in the observation that DF models encounter challenges in generating high-frequency (HF) signals due to overfitting to human speech. This phenomenon arises primarily from the irrelevant nature of HF signals, such as background bird noise, during training. DF models prioritize learning speech waveforms, neglecting uncorrelated high-frequency noise. Our approach is complementary to advances in generative models and instead pertains to authenticity-critical applications such as banking and election campaigns.\\
\underline{\textit{Literature Review:}}
Various methods have been explored to ensure the speech authenticity, as in Fig.\ref{fig:pipeline}. Anonymization is commonly employed to obscure speech identity, and it significantly degrades speech naturalness and intelligibility \cite{cohen2019voice,
wu20p_interspeech}. Watermarking stands out for embedding information within audio signals, facilitating post-processing regeneration \cite{agaskar22_interspeech, chen20i_interspeech, zhang2024v2a}. Despite its utility, the ease of the watermark extraction poses security threats. Besides, the objectives of anonymization (hiding user identity) and watermarking (maintaining data copyright) are distinct from our DF countermeasure (securing user identity). Signal processing methods offer cost-effective signatures but suffer from reversibility, which allows the extraction of the original signature since both the original speech and its signed version are public \cite{campi2021machine, guo2017voice}. ML classifiers have emerged to discern between original and cloned signals \cite{kawa23b_interspeech, mun23towards,liu23v_interspeech, wang2023robust}. However, a generative DF model is trained to deceive its discriminators, analogous to these classifiers \cite{creswell2018generative}. Hence, their performance is limited by the DF models' discriminators.\\
\underline{\textit{Principal Contributions:}} In light of prior work, our contributions are four-fold. First, we systematically analyze the impact of DF models on the frequency band, highlighting a significant energy reduction in the HF regime, as empirically demonstrated in Fig. \ref{fig:empirical_result}. Second, we introduce a secure signature method leveraging this observation to detect unauthorized DF attacks in voice-authenticated systems. Our method, depicted in Fig.\ref{fig:pipeline}, surpasses the performance of existing state-of-the-art ML-based detection mechanisms. Third, we explore the integration of Differential Privacy (DP) techniques to safeguard private key integrity, mitigating the risk of unauthorized access. Finally, we open-source our code\footnotemark[1], facilitating further research in this area.
\section{Methodology}\label{sec:methodology}
SecureSpectra comprises three key components: a generative model for DF attacks, a signature model, and a verification model to detect the signature's presence, as shown in Fig.\ref{fig:pipeline}.\\\\
\underline{\textbf{Adversary's Threat Model:}}
We now describe our assumptions on the adversary's threat model and how each component in Fig.\ref{fig:pipeline} mitigates different types of attacks. First, the adversary can train a DF model to create a forged audio. However, our signature module (green) effectively thwarts such attempts. Second, a sophisticated adversary can steal the private signature model's weights. However, we combat this by making the users sign their audio with their private keys (red). Finally, authenticated adversaries with a signature model can attempt to reverse engineer private keys from the verifier model. To combat this, we add DP to private keys during training to effectively prevent such reverse engineering attacks. We now describe each of these threats and their defense in turn.\\\\
\underline{\textbf{Generative DeepFake Model:}}\label{sec:DF}
Let $\{a_i\}_{i=1}^N$ be $N$ unique samples drawn from the data distribution $\mathbf{D}$. An adversary trains a GAN on these samples to create synthetic audio $\Tilde{a}_i^\mathcal{T}$ that resembles the original $a_i$ and vocalizes the provided text $\mathcal{T}$ from a text corpus $\mathbf{T}$. We focus on GANs for notational convenience, though the concepts apply to any generative models, such as diffusion models \cite{ho2020denoising} and VAEs \cite{kingma2013auto}. A GAN comprises a Generator $\mathcal{G}$ and a Discriminator $\mathcal{D}$. Generator $\mathcal{G}(a_i,\mathcal{T};\theta_\mathcal{G})$, parameterized by $\theta_\mathcal{G}$, takes a reference audio $a_i$ and a transcript $\mathcal{T}$ and generates a cloned audio $\widetilde{a}_i^\mathcal{T}$ corresponding to the transcript. Discriminator $\mathcal{D}(a;\theta_\mathcal{D})$, parameterized by $\theta_\mathcal{D}$, takes audio data $a$ and predicts whether it is synthetic $\Tilde{a}_i$, or not $a_i$. Formally, the problem is expressed as:
\begin{align} \small
    \min_{\theta_\mathcal{G}} \max_{\theta_\mathcal{D}} &\mathbb{E}_{a\sim \mathbf{D}} [\log \mathcal{D}(a;\theta_\mathcal{D})] \notag \\
    &+\mathbb{E}_{a,\mathcal{T} \sim \mathbf{D},\mathbf{T}} [\log(1 - \mathcal{D}(\mathcal{G}(a,\mathcal{T};\theta_\mathcal{G});\theta_\mathcal{D}))].
\end{align}
The objective is to train the generator parameters $\theta_\mathcal{G}$ to produce synthetic audio that is indistinguishable from real audio by deceiving the discriminator, while the discriminator parameters $\theta_\mathcal{D}$ are trained to distinguish original samples from clones. \\\\
\underline{\textbf{Signature Model:}} Our main goal is to sign the audio $a_i$ such that its clone $\widetilde{a}_i^\mathcal{T} = \mathcal{G}(a_i^*,\mathcal{T};\theta_\mathcal{G})$ lacks the signature. We introduce a signature module $\mathcal{S}$ that operates on an audio $a_i$ and a private key $\kappa_i$ to generate a signed version of the audio $a^*_i$. Each user has a distinct private key $\kappa_i$ to sign the audio samples: $a_i^* = S(a_i, \kappa_i; \theta_\mathcal{S}).$ The parameters $\theta_\mathcal{S}$ are optimized to minimize the $\ell_1$ norm of the original and signed audio samples. The $\ell_1$ norm encourages sparsity in the spectrum with minor changes, preserving HF quality \cite{candes2008enhancing}. The signature module's loss is formulated as:
\begin{equation} \small
    \mathcal{L}_\mathcal{S}:= \frac{1}{N}\sum_{i=1}^N||a_i-S(a_i,\kappa_i;\theta_\mathcal{S})||_1.
\end{equation}
We fuse the private key with the audio representation at the deepest layer of the U-Net architecture \cite{ronneberger2015u}. Such integration ensures that the private key is intricately woven into the utterance, thereby complicating any attempts at extraction without compromising the audio quality. This mechanism secures the keys and ensures the signature embedding is inherently resilient to reverse engineering. This resilience is further bolstered by the signature model's confidentiality and DP (Sec.~\ref{sec:DP}). Deciphering the embedded signature without explicit knowledge of the signature model and private key becomes extremely challenging. This layered approach to security ensures the integrity of audio signatures, providing a robust defense against unauthorized access and manipulation.\\
\begin{figure}[t]
    \centering
    \includegraphics[width=0.8\linewidth]{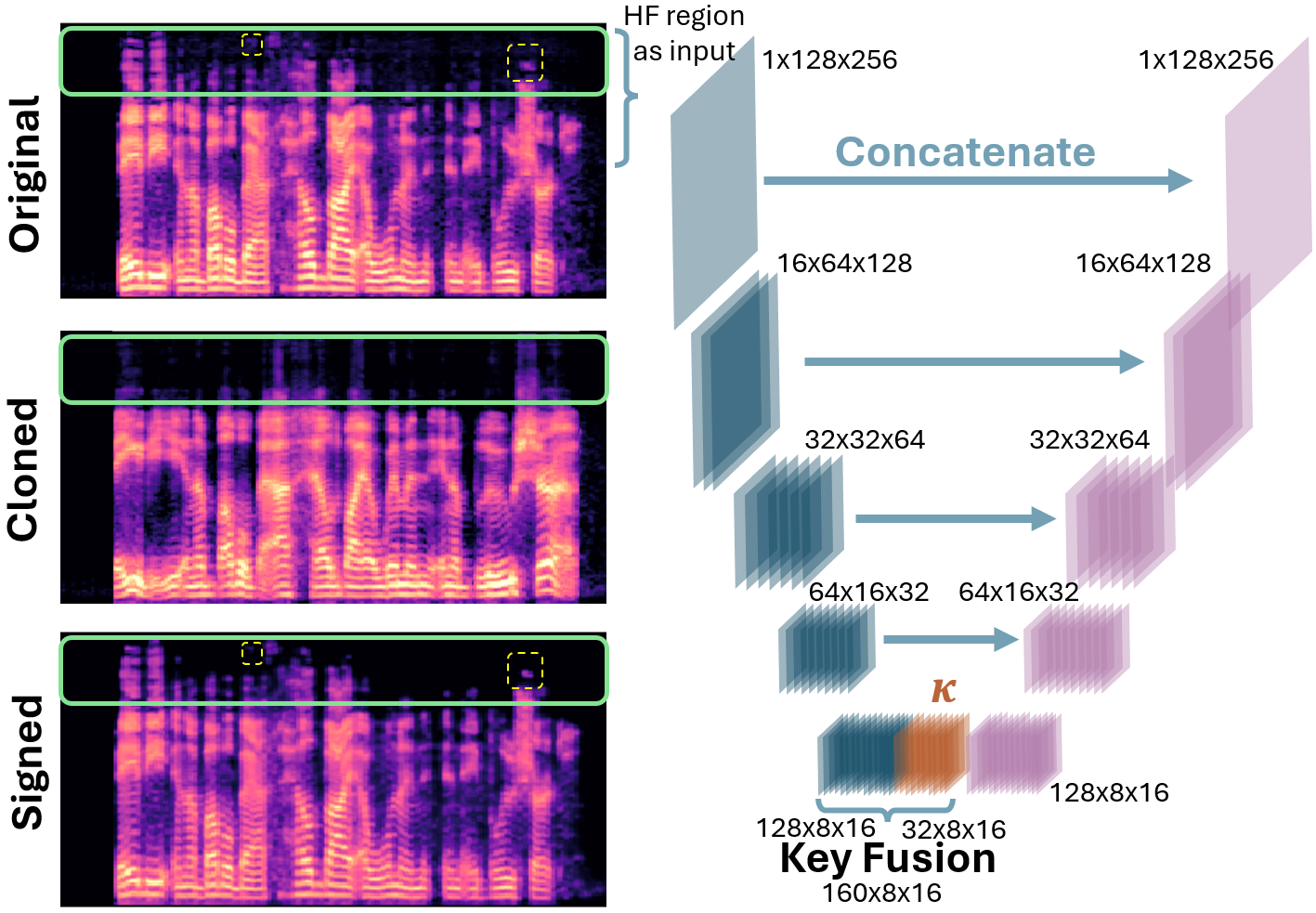}
    \caption{\textbf{Key Observation for High-Frequency Regime:} By comparing the spectrograms of the original audio (top) and its cloned version (middle) for the same transcript, we observe a distinct absence of HF content (green) in the DF audio. 
    This discrepancy arises from the bias of DF models toward mimicking user speech, which predominantly emphasizes lower-frequency regions. A U-net (right) signs the audio (bottom) with unrecognizable slight modifications (yellow) in the HF regime.}
    \label{fig:microscale_result}
\end{figure}\\
\underline{\textbf{Verification Model:}} Following the design of the signature model, our next logical step is to introduce a verification model $\phi$ designed to classify audio signals as signed or not. It is a \textit{public} model and reveals ``only the existence of the signature''. The model $\phi$ outputs a binary prediction $\hat{y}$ for audio $a_i$ through $\hat{y} = \phi(a_i;\theta_\phi)$, where $\theta_\phi$ represents the parameters trained to minimize the verifier loss $\mathcal{L}_\phi$ before public release. For given unsigned audio with ground truth $(a_i,y_i)$ and signed audio with ground truth $(a_i^*,y_i^*)$, the verifier loss $\mathcal{L}_\phi$ is defined as:\\
\begin{equation}
    \mathcal{L}_\phi:=-\frac{1}{N}\sum_{i=1}^N y_i\log(\phi(a_i;\theta_\phi))+(1-y_i^*)\log(1-\phi(a_i^*;\theta_\phi)),
\end{equation}
which is the extended binary cross-entropy loss for our setting.\\\\
\underline{\textbf{Joint Training:}} After designing the models, we ensure that the signature and verification modules complement each other's functions. Achieving both hidden signatures and accurate verification requires the signature model to integrate verifier gradients during training. Thus, both models learn from each other by minimizing the joint loss shown as:\begin{equation}
    \min_{\theta_\phi,\theta_\mathcal{S}} \mathcal{L}_\mathcal{S}+\mathcal{L}_\phi. \label{eqn:jointloss}
\end{equation}The training begins with a forward pass through the signature model to compute its loss. Then, both the signed and original audio samples undergo a forward pass in the verification model to calculate the verifier loss. These two losses are then aggregated, as shown in Eq. \ref{eqn:jointloss}, for optimization. Subsequently, the gradients propagate backward through both models, with particular attention to the verification model.  To calibrate the potential instabilities arising from the verification module and affecting the signature module, the learning rate for the verifier model is set to one-tenth of that for the signature model.\\\\
\underline{\textbf{Guarded Inference:}}\label{sec:guardedinf}
Following the training phase, the verification model becomes publicly accessible for authenticity and signature checks, while the signature model remains confidential to prevent unauthorized signature replication. This approach safeguards against adversarial attempts to create counterfeit signatures because it requires both the signature model and private keys to generate valid signatures. Only authorized users with private keys can embed signatures, maintaining the process's security and privacy. This dichotomy effectively protects digital audio identities from unauthorized access or manipulation.\\\\
\underline{\textbf{Differential Privacy (DP) on Private Keys:}}~After~the~defense against external threats, we now address insider risks with DP. Authorized malicious users or exposure of the signature model allows adversaries to reverse engineer the system and deduce private keys used in model training \cite{fredrikson2015model}. To address this, we introduce precisely calibrated DP noise to the private keys during the signature model's training. This ensures that the model can embed signatures without incorporating identifiable information about individual keys into its parameters. This prevents adversaries from extracting private keys when they have the signature model. Formally, private keys $\kappa$ are preserved by adding DP noise $\eta$ sampled as:\begin{equation}
        \eta \sim \frac{1}{2b} \exp\left(-\frac{|\eta|}{b}\right),
\end{equation}
where the noise scale $b$ is the ratio $\Delta K /\epsilon$. Here, $\Delta K$ represents the global sensitivity, indicating the maximal cosine distance between any two private keys $\kappa$.~$\epsilon$ denotes the privacy loss. It configures the trade-off between key protection and model accuracy and ensures adherence to the $\epsilon$-DP standard \cite{dwork2006differential}. This standard mandates that the alteration of a single element in a private key, resulting in a transition from $\kappa$ to $\kappa^\prime$, should not significantly affect the probability distribution of the output of the model $\mathcal{S}$, as in $\Pr[\mathcal{S}(\kappa) \in R] \leq e^\epsilon \times \Pr[\mathcal{S}(\kappa^\prime) \in R]$, where $R$ represents any arbitrary subset of outcomes.
\label{sec:DP}
Each operation in the pipeline enhances the integrity of signed audio, addressing DF threats and privacy concerns in voice-based applications. Each signature is intricately tied to the utterance and private key, adding extra robustness.
\section{Experimental Setup} \label{sec:experiments}
In this section, we detail the experimental setup, outlining the datasets, models, hyperparameters, and evaluation metrics.\\
\underline{\textbf{Datasets:}} To demonstrate the efficacy of our proposed pipeline, we utilize three widely recognized speech datasets: \textbf{Mozilla Common Voice} \cite{ardila2019common}, \textbf{LibriSpeech} \cite{panayotov2015librispeech}, and \textbf{VoxCeleb} \cite{nagrani2017voxceleb}. The ASVspoof2021 \cite{yamagishi2021asvspoof} is unsuitable for our analysis as it lacks regenerable DFs from signed audio to compare its effects.\\
\underline{\textbf{DF Models:}} We evaluated the resilience of SecureSpectra and the effects of DF models on spectrograms. We selected three state-of-the-art GAN models for this purpose: \textbf{Coqui.ai TTS} \cite{coqui}, \textbf{OpenVoice} \cite{openvoice}, and \textbf{WhisperSpeech} \cite{whisperspeech}. These models were chosen for their ability to synthesize speech from text inputs using ``reference audio,'' making them ideal for examining the impact of DF on the authenticity of signed audio. Using these models, we cloned audio from speech datasets to vocalize texts by Camus, Dickens, Orwell, and Thoreau. This process created cloned datasets that showcased not only the HF synthesis capabilities of the DF models but also our verification system's ability to detect audio authenticity. \\
\underline{\textbf{Benchmarks:}} Our evaluation of SecureSpectra encompasses three distinct configurations: Verification Only, Signature + Verification, and Signature + Verification with DP noise. The \textbf{Verification Only} serves as a baseline, demonstrating detection performance without our signature embedding technique. It highlights the improvement done by our signature mechanism. The core configuration, \textbf{Signature + Verification}, demonstrates robustness against DF attacks, achieved through the joint training of the signature and verification modules. Introducing \textbf{DP into the Signature + Verification} configuration prevents reverse engineering threats and balances performance-privacy trade-offs in trustless settings. Furthermore, SecureSpectra's performance is compared with two state-of-the-art anti-spoofing solutions from INTERSPEECH 2023. The first method, \textbf{Whisper Based} \cite{kawa23b_interspeech}, uses the Whisper \cite{radford2023robust} features of audio to detect DF attacks with an ML classifier. The second, \textbf{SASV2-Net} \cite{mun23towards}, utilizes a multi-stage training approach that transfers information from speaker verification in the VoxCeleb dataset to DF detection on the ASVspoof2019 dataset, showcasing orthogonal information leverage similar to our approach in some aspects.\\
\underline{\textbf{Evaluation Metric:}} To compare the benchmarks, we assess the test accuracies of classification models on the recordings from 100 individuals separately. Each individual's test set comprises cloned and original samples with equal density. Also, we evaluate the benchmark Equal Error Rates (EERs) on each dataset.\\
\underline{\textbf{Signatures:}} Each key $\kappa$, consisting of 32 binary digits, can be viewed as vectors. To gauge the global sensitivity $\Delta K$ of these keys, we compute the maximum cosine distance between them. Subsequently, we introduce DP noise $\eta$ with $\epsilon=30$. The noise is injected during training ($\kappa+\eta$) to ensure that the model learns from a distribution of private keys rather than individual ones.\\ 
\underline{\textbf{Model Architectures:}} The signature model, shown in Fig.\ref{fig:microscale_result}, employs a U-Net tailored for processing the HF spectrograms of audio. It consists of a 5-layer CNN encoder that projects the input. At the deepest layer of the encoder, the private key $\kappa$ is concatenated with the latest feature map and convolved to maintain the original feature map size. This augmented feature map then undergoes decoding through four layers, each receiving a skip connection from the encoder at the corresponding level. For verification, a 7-layer CNN with kernel size 3 condenses the input features by doubling the channel size in each layer, ultimately leading to a fully connected layer that outputs a binary prediction on the presence of $\kappa$ without disclosing any other information. More model details are available in our repository\footnotemark[1].\\
\underline{\textbf{Hyperparameters:}} Both models are initialized using Xavier initialization \cite{glorot2010understanding}, and the training is performed using the Adam optimizer \cite{kingma2014adam} with an initial learning rate of 2e-4 for the signature model and 2e-5 for the verification model. Training is carried out to minimize total validation loss, with early stopping after 10 rounds of no improvement. The experiments were executed with 8 NVIDIA RTX A5000 GPUs (8x24 GB RAM).

\section{Results} \label{sec:results}

\begin{figure}[t]
    \centering
    \includegraphics[width=0.9\linewidth]{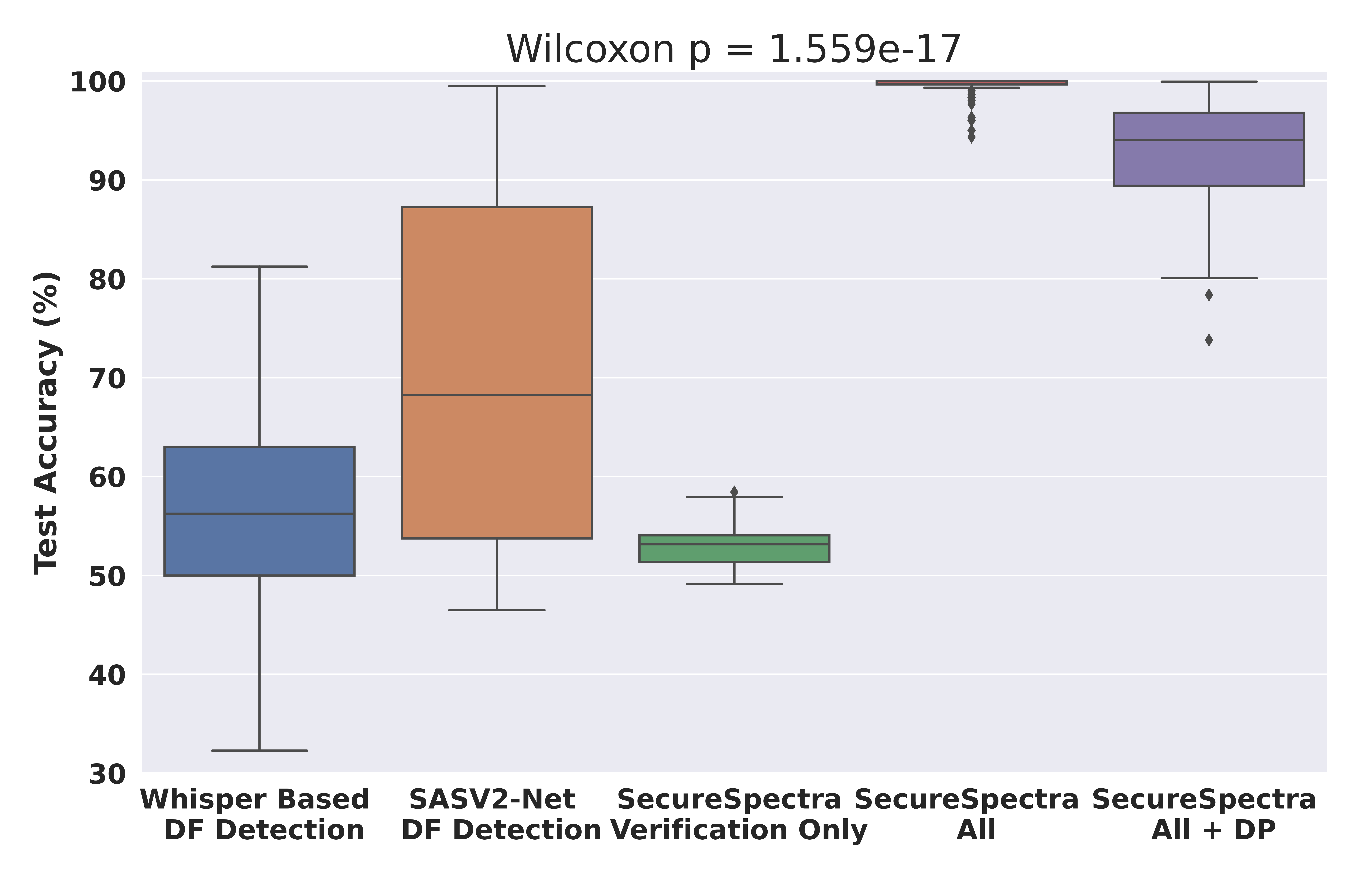}
    \caption{\textbf{User-Level Performance Across Benchmarks:} We evaluate the DF detection benchmark test accuracies across 100 distinct users with 200 audio samples each (100 original, 100 cloned). The orange and blue box plots show the accuracies of the two recent works. The green box plot provides a baseline of our pipeline without signature embedding. The purple and red box plots show the performance of our approach with and without DP noise, respectively. \textbf{Our method, particularly with signature embedding, surpasses existing models}, enhancing verification-only accuracy by 81\% and outperforming comparative works by 71\% and 42\%. DP noise adds additional security with a marginal decrease in accuracy by 4\%.}
    \label{fig:macroscaleresults}
\end{figure}

Our empirical analysis (Fig.\ref{fig:empirical_result}) demonstrates that DF models tend to overlook HF patterns while generating cloned speech. Furthermore, in Fig.\ref{fig:macroscaleresults} and Table \ref{table:EERtable}, we demonstrate that SecureSpectra performs better than recent work in the literature. Additionally, we show that adding DP noise reduces performance slightly with the benefit of a more secure pipeline.\\ 
\textit{\textbf{How does orthogonal information affect DF detections?}}\\
Our analysis suggests that DF generators, designed to deceive their intrinsic GAN discriminators, similar to our Verification Only module, can be effectively countered by leveraging orthogonal information. This is evidenced by the performance improvements observed when transitioning from a conventional CNN detection module (green) to methods employing auxiliary information. Specifically, Whisper Based (blue) and SASV2-Net (orange) leverage transcription and speaker verification information, respectively. Unlike these approaches, which rely on information correlated with the audio's content, SecureSpectra embeds completely orthogonal information, the $\kappa$ signature, into the audio (red and purple). This strategy yields a robust mechanism for DF detection across all user scenarios.
\underline{\textbf{Limitations:}} SecureSpectra's scalability is linked to the length of private keys. As user numbers grow, the requisite key length increases, potentially complicating the model's training process. Also, our evaluation excludes the audio channel noise.

\begin{table}
\caption{\textbf{The Benchmark $\%$ EERs ($\downarrow$) on Individual Datasets:} SecureSpectra dramatically reduces EER across all datasets.}
\centering
\begin{tabular}{lccc}
 Benchmarks & CV \cite{ardila2019common} & LS \cite{panayotov2015librispeech} & VC \cite{nagrani2017voxceleb} \\ \hline
Whisper \cite{kawa23b_interspeech} & 41.6 & 36.2 &  36.8 \\
SASV-2 \cite{mun23towards} & 4.01 & 12.2 & 3.75 \\
Verification Only & 48.3 & 43.6 & 46.0 \\
\textbf{SecureSpectra} & 1.50 & 1.10 & 1.36 \\
\textbf{SecureSpectra + DP} & 2.96 & 2.74 & 2.83
\end{tabular}
\label{table:EERtable}
\end{table}

\section{Conclusion}\label{sec:conclusion}
This paper introduces SecureSpectra, a robust method for protecting audio from cloning attacks using HF signatures that DF models cannot accurately reproduce. Through empirical analysis, we show the inability of DF models to mimic HF signals, thereby significantly increasing detection accuracy by embedding signatures. We open-source SecureSpectra\footnotemark[1] to support ongoing research. As DF technologies advance, SecureSpectra offers robust protection for digital identity. Moving forward, we aim to enhance SecureSpectra by advancing the verification module through multitask learning with speaker verification. This approach ensures that users can only sign their personal audio, strengthening both robustness and security.

\section{Acknowledgements}
This material received support from the National Science Foundation under grant no.2148186 and is further supported by funding provided by federal agencies and industry partners as specified in the Resilient \& Intelligent NextG Systems (RINGS) program. This article solely reflects the opinions and conclusions of its authors and does not represent the views of any sponsor.

\bibliographystyle{IEEEtran}
\bibliography{mybib}

\end{document}